\documentclass[3p,times,procedia]{elsarticle}
\usepackage{nupha_ecrc}

\volume{00}

\firstpage{1}

\journalname{Nuclear Physics A}

\runauth{B. Sahlmuller}

\jid{nupha}

\jnltitlelogo{Nuclear Physics A}

\usepackage{amssymb}
\usepackage{wrapfig}

\usepackage[figuresright]{rotating}

\begin{document}

\begin{frontmatter}



\dochead{}

\title{Direct photon measurement in Pb-Pb collisions\hspace{4cm} at $\sqrt{s_{\rm NN}}$ = 2.76 TeV with ALICE}


\author{Baldo Sahlm\"uller for the ALICE Collaboration}

\address{IKF, Goethe University Frankfurt, Max-von-Laue-Str. 1, 60438 Frankfurt, Germany}

\begin{abstract}

The ALICE experiment has measured the direct photon spectra in Pb-Pb collisions at $\sqrt{s_{NN}} = 2.76$~TeV for three different centrality selections. The measurement was performed emplying a method utilizing conversion of photons into $e^+e^-$ pairs in the detector material, and a method using the PHOS calorimeter. The two measurements were combined in order to measure direct photons over a broad transverse momentum range of $0.9 < p_{\rm T} < 14$ GeV/$c$. A direct photon signal was observed in the thermal region $0.9 < p_{\rm T} < 2.1$ GeV/$c$, with a significance of $2.6\sigma$ in the 20\% most central collisions. The corresponding direct photon spectrum can be described by an exponential with an inverse slope parameter of $(304 \pm 11^{\rm stat} \pm 40^{\rm syst})$ MeV, without subtracting the hard scattering component, which is somewhat larger than the inverse slope parameter of the direct photon spectrum at RHIC energies. Within uncertainties, the data can be described by different models for photon production in heavy-ion collisions. These proceedings provide a summary of the results published in~\cite{Adam:2015lda} and~\cite{ALICE-PUBLIC-2015-007}.
\end{abstract}


\end{frontmatter}


\section{Introduction}
\label{sec:into}

Direct photons are produced in all stages of ultrarelativistic heavy-ion collisions that are performed with lead nuclei at a center-of-mass energy of $\sqrt{s_{\rm NN}} = 2.76$~TeV at the LHC. Since they do not interact strongly they leave the dense and hot nuclear medium produced in the collision unscathed. Direct photons are thus believed to provide information about all stages of the collisions~\cite{Turbide:2003si}. At high $p_{\rm T}$, direct photon production is dominated by hard scattering processes and can thus be used to study the validity of binary or $N_{\rm coll}$ scaling of high $p_{\rm T}$ particle production. At low $p_{\rm T}$, direct photons are produced in the quark-gluon plasma and the hadron gas phase. Since these photons are from a thermal source, their spectrum follows approximately an exponential distribution and contain information about the initial temperature of the medium and its subsequent evolution. However, the direct photon spectrum at low $p_{\rm T}$ is a sum of contributions from all stages of the space-time evolution of the medium, hence the interpretation of the measured photon spectrum is not straightforward and based on assumption on the initial temperature, the expansion of the medium, and blue-shift due to radial flow~\cite{Shen:2013vja}.\\

An earlier direct photon measurement in heavy-ion collisions was performed by the PHENIX experiment, using internal conversion of photons~\cite{Adare:2008ab}. The measurement confirmed the assumption of binary scaling between the high $p_{\rm T}$ $(\gtrsim 5$ GeV/$c)$ direct photon production in pp and Au-Au collisions at $\sqrt{s_{\rm NN}} = 200$ GeV. At low transverse momenta ($1 \lesssim p_{\rm T} \lesssim 4$ GeV/$c$), they measured an excess of photons over the expectation of $N_{\rm coll}$ scaled pp data that could be parameterized by an exponential function with an inverse slope parameter of $T_{\rm eff} = 221 \pm 19^{\rm stat} \pm 19^{\rm syst}$ MeV. This measurement was later confirmed by a measurement using conversion in the detector material where the slope was found to be $T_{\rm eff} = 239 \pm 25^{\rm stat} \pm 7^{\rm syst}$ MeV~\cite{Adare:2014fwh}.

At the LHC, isolated direct photons at large transverse momentum $22 < p_{\rm T} < 280$ GeV/$c$ and $20 < p_{\rm T} < 80$ GeV/$c$ were measured by ATLAS~\cite{Aad:2015lcb} and CMS~\cite{Chatrchyan:2012vq}, respectively. These measurements confirmed the binary scaling of direct photon production at high $p_{\rm T}$.

\section{Measurement}
\label{sec:measure}

\begin{wrapfigure}{l}{0.45\textwidth}
  \begin{center}
\includegraphics[width=0.45\textwidth]{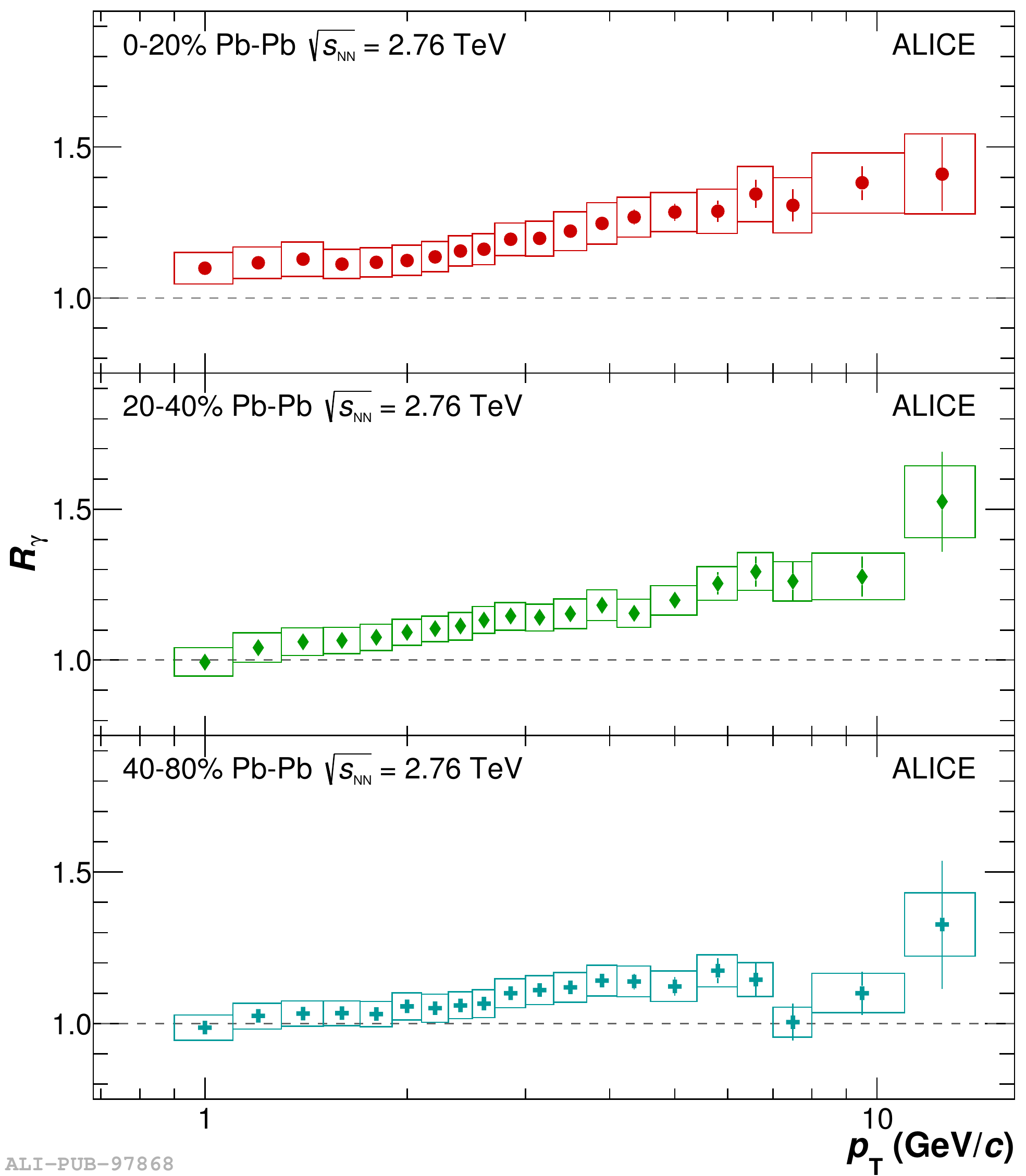}
\end{center}
\caption{\label{fig:rgamma} Combined PCM and PHOS double ratio $R_{\gamma} = (\gamma_{\rm inclusive}/\pi^{0}_{\rm param}) / (\gamma_{\rm decay}/\pi^{0}_{\rm param})$ for three different centrality selections 0-20\%, 20-40\%, and 40-80\%~\cite{ALICE-PUBLIC-2015-007}.}
\end{wrapfigure}

The ALICE detector has various capabilities to measure photons and their energy. For this analysis, photons were measured in two independent analyses, using the ALICE tracking and PID capabilities by reconstructing secondary vertices of $e^+ e^-$ pairs from photons converted in the detector material, and using the ALICE Photon Spectrometer (PHOS), a highly segmented electromagnetic calorimeter. The details of the measurement have been published in~\cite{Adam:2015lda}.

The Photon Conversion Method (PCM) benefits from an excellent energy resolution at low $p_{\rm T}$, with a $\pi^{0}$ peak width of $\sigma_{\pi^{0}} \approx 3$ MeV/$c^2$. A high purity of the photon sample of $>90\%$ in central collisions and $>98\%$ in peripheral collisions was obtained by using cuts on the decay topology and on electron track properties. Due to the conversion probability of $\sim 8.5\%$, the measurement is limited by statistics at high $p_{\rm T}$, however it benefits from a large acceptance of $2\pi$ in azimuth and $|\eta| < 0.9$.

The PHOS calorimeter measures photons that leave electromagnetic showers in the detector material. It covers an azimuthal range of $\sim 60$ degrees and a pseudorapidity range of $|\eta|<0.13$. Photons are measured with a high energy resolution that results in a $\pi^{0}$ peak width of 
$\sigma_{\pi^{0}} \approx 7$ MeV/$c^2$ at low $p_{\rm T}$. Photons in the PHOS are selected by cuts on energy, size, and dispersion of calorimeter clusters, resulting in a high purity of the photon sample of $96-99 \%$.\\

Both analyses use a statistical method to extract the direct photon signal from the data. First, inclusive photon ($\gamma_{\rm inclusive}$) spectra are measured using both analysis techniques. In a next step, the decay photon ($\gamma_{\rm decay}$) spectra are calculated, based on the measured $\pi^{0}$ spectrum, for each method individually. The decay photons from other mesons such as $\eta$, $\omega$, $\eta'$ are derived from the $\pi^{0}$ spectrum, using $m_{\rm T}$ scaling for most mesons. Only for the $\eta$ meson which contributes to $\sim 10-12\%$ of the decay photons, a different approach was used to account for a possible flow contribution. Hence, an alternative scaling was considered, using the measured shape of the $K^{0}_{S}$ spectrum as a proxy for the shape of the $\eta$ spectrum. For the calculation of the decay photons, an average of $m_{\rm T}$ scaling and the $K^{0}_{S}$ shape was used.\\

In order to reduce some uncertainties, the decay photon contribution was subtracted by calculating the so-called double ratio
$R_{\gamma} = (\gamma_{\rm inclusive}/\pi^{0}_{\rm param}) / (\gamma_{\rm decay}/\pi^{0}_{\rm param})$, and then calculating the direct photon spectrum as $\gamma_{\rm direct} = (1 - 1/R_{\gamma})\cdot \gamma_{\rm inclusive}$. The measured $\pi^{0}$ spectrum was parameterized ($\pi^{0}_{\rm param}$) in order to calculate the decay photon spectrum, this parameterization was also used in the double ratio.

The systematic uncertainties of the inclusive photon spectrum in the PCM analysis are dominated by the uncertainties on the material budget of the detector $(\sim 4.5\%)$, and on the photon selection criteria (up to $\sim 4\%$ at low $p_{\rm T}$). The main uncertainties of the PHOS measurement are the uncertainties on the energy scale of the detector $(\sim 9.5\%)$, and on the reconstruction efficiency calculation ($\sim 3\%$ in 0-20\% central collisions).

The consistency of the two analyses was checked at different levels, using pseudo experiments to calculate test statistics and subsequently a level of agreement between the analyses. The agreement was found to be 1.2 standard deviations for the inclusive photon spectra, and 0.4 standard deviations for $R_{\gamma}$. Hence, the results of the two analyses were combined into a single direct photon spectrum. More details on the analyses and on the comparisons can be found in~\cite{Adam:2015lda}.

\begin{figure}
\begin{minipage}{0.5\textwidth}
\includegraphics[width=0.99\textwidth]{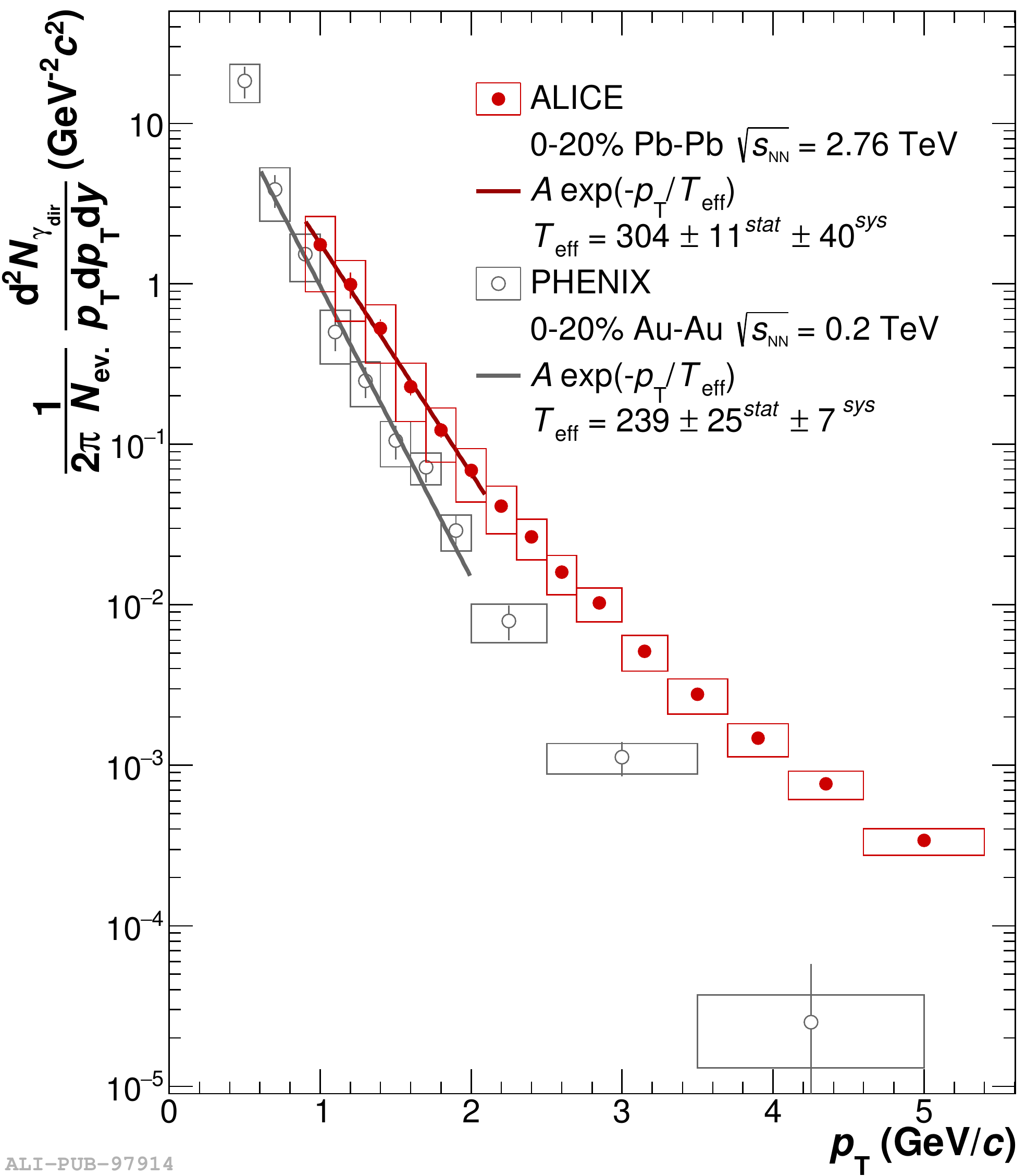}
\end{minipage}
\begin{minipage}{0.5\textwidth}
\includegraphics[width=0.99\textwidth]{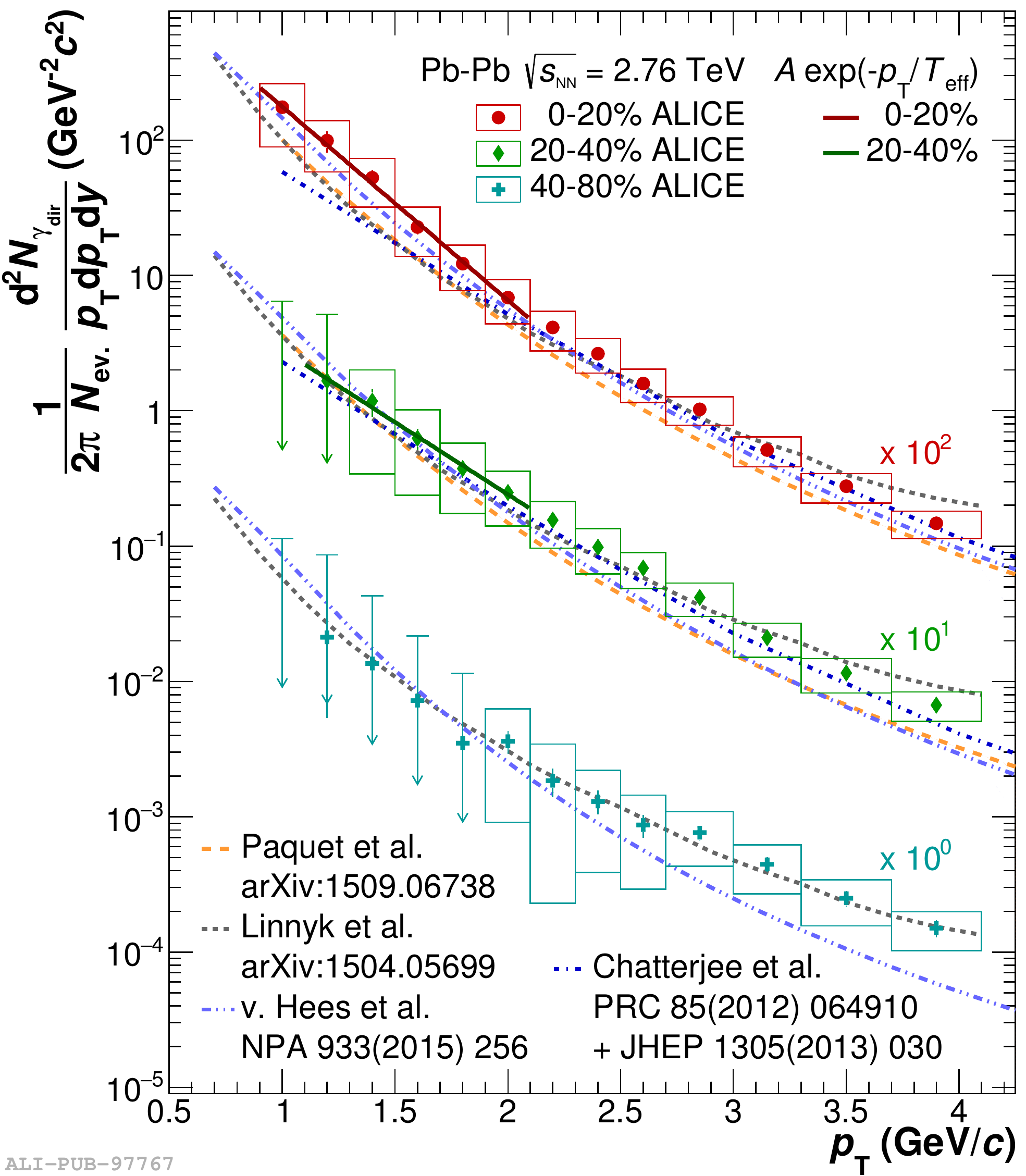}
\end{minipage}
\caption{\label{fig:spectra} Left: Direct photon spectrum for the 0-20\% most central events in Pb-Pb collisions at $\sqrt{s_{\rm NN}} = 2.76$ TeV and in Au-Au collisions at  $\sqrt{s_{\rm NN}} = 0.2$ TeV~\cite{Adare:2014fwh}. An exponential fit is shown for both spectra, covering the range $0.9 < p_{\rm T}Ê< 2.1$ GeV/$c$ and $0.6 < p_{\rm T}Ê< 2.0$ GeV/$c$, respectively~\cite{ALICE-PUBLIC-2015-007}. Right: Direct photon spectra for three different centrality classes (0-20\%, 20-40\%, and 40-80\%), compared to theoretical calculations from different models~\cite{Paquet:2015lta, vanHees:2014ida, Chatterjee:2012dn, Linnyk:2015tha}. The 0-20\% centrality class was scaled by a factor of 100, the 20-40\% class by a factor of 10. In addition, the 0-20\% and the 20-40\% centrality classes are fit with an exponential function~\cite{Adam:2015lda}.}
\end{figure}

\section{Results}
\label{sec:results}
The combined double ratio $R_{\gamma}$ is shown in Fig.~\ref{fig:rgamma}, for three different centrality classes. A clear excess is observed at large $p_{\rm T}$ in all centrality classes. At low $p_{\rm T}$, an excess is observed with a significance of $2.6\sigma$ in the 0-20\% most central event. The significance of the excess was quantified with the same method as the consistency check between the two analyses, using $R_{\gamma} = 1$ as null hypothesis.

The direct photon spectrum for the 0-20\% most central Pb-Pb events at $\sqrt{s_{\rm NN}} = 2.76$~TeV is shown in the leftt panel of Fig.~\ref{fig:spectra}, together with the spectrum for the same centrality selection in Au-Au collisions at $\sqrt{s_{\rm NN}} = 0.2$~TeV~\cite{Adare:2014fwh}. Both spectra are fit with an exponential function at low $p_{\rm T}$ in order to extract the respective slope parameters. The fit to the spectrum from~\cite{Adare:2014fwh} gives a slope parameter of $T_{\rm eff} = 239 \pm 25^{\rm stat} \pm 7^{\rm syst}$ MeV, while the ALICE data were found to have a larger slope parameter of $T_{\rm eff} = 304 \pm 11^{\rm stat} \pm 40^{\rm syst}$ MeV which is consistent with a larger initial temperature of the medium at the larger centre-of-mass energy. Note that for the fit to the ALICE data, the pQCD contribution was not subtracted to obtain the values given here while PHENIX subtracted their measured $N_{\rm coll}$ scaled pp spectrum from the Au+Au data.

The direct photon spectra for all three centrality classes are shown in the right panel of Fig.~\ref{fig:spectra}, in comparison to various theoretical models~\cite{Paquet:2015lta, vanHees:2014ida, Chatterjee:2012dn, Linnyk:2015tha}. All models include photons  from hard scattering processes calculated by pQCD. The models from~\cite{Paquet:2015lta, vanHees:2014ida, Chatterjee:2012dn} incorporate hydrodynamic calculations with different descriptions of initial conditions, different formation times $\tau_{0}$ and different initial temperatures $T_{\rm init}$. The model from~\cite{Linnyk:2015tha} is based on a transport approach and a microscopic description of the evolution. All models agree with the data within uncertainties.\\ 

\begin{figure}
\begin{minipage}{0.5\textwidth}
\includegraphics[width=0.99\textwidth]{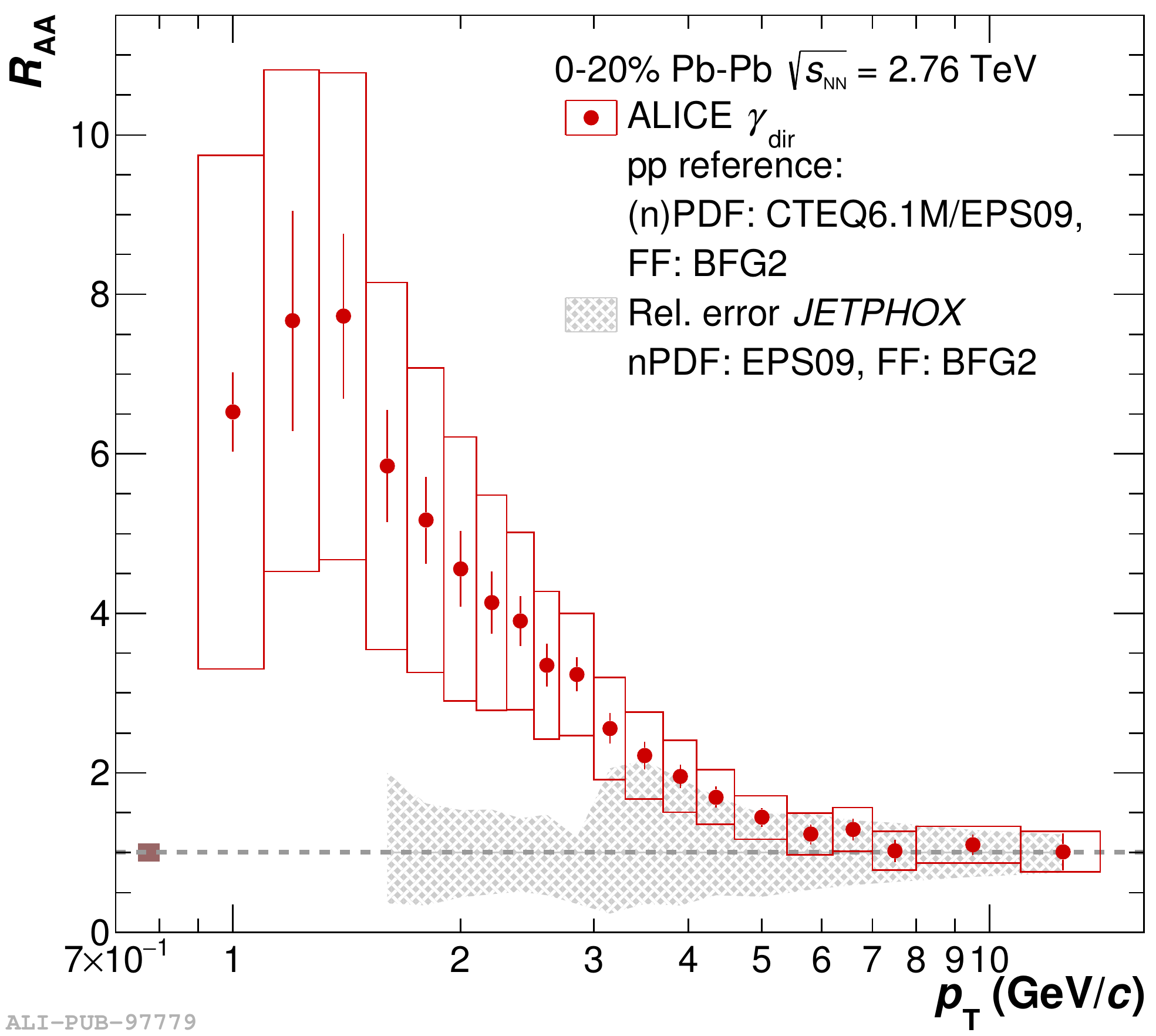}
\end{minipage}
\begin{minipage}{0.5\textwidth}
\includegraphics[width=0.99\textwidth]{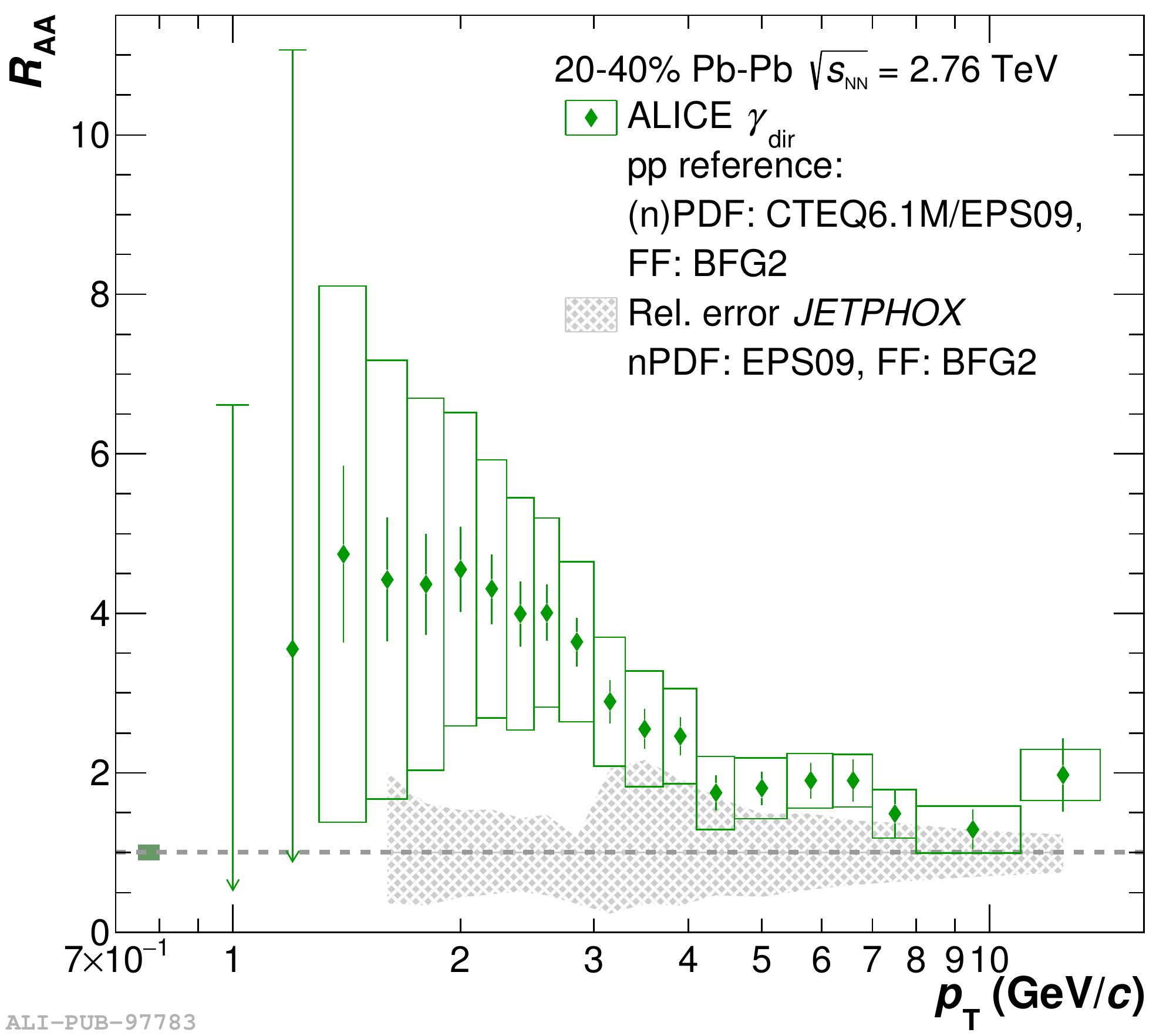}
\end{minipage}
\caption{\label{fig:raa} Nuclear modification factor $R_{\rm AA}$ for the 0-20\% (left) and 20-40\% (right) centrality classes~\cite{ALICE-PUBLIC-2015-007}. The pQCD calculation from~\cite{Paquet:2015lta} was used as pp reference.}
\end{figure}

The nuclear modification factor $R_{\rm AA}$ can be used to quantify nuclear effects in heavy-ion collisions, it was calculated for direct photons using the pQCD calculation from~\cite{Paquet:2015lta} as pp reference. It is shown in Fig.~\ref{fig:raa} for the 0-20\% and 20-40\% centrality classes. $R_{\rm AA}$ is a different way to show the strong enhancement of the direct photon production at low $p_{\rm T}$ over the expectation from $N_{\rm coll}$ scaled pp collisions, as well as the validity of binary scaling at high $p_{\rm T}$.

\section{Summary}

In summary, ALICE has measured the direct photon spectra and $R_{\rm AA}$ for different centrality selections in Pb-Pb collisions at $\sqrt{s_{\rm NN}} = 2.76$~TeV, using two independent analysis methods~\cite{Adam:2015lda}. The inverse slope parameter of an exponential function fit to the spectrum was extracted and found to be $T_{\rm eff} = 304 \pm 11^{\rm stat} \pm 40^{\rm syst}$ MeV in the 0-20\% centrality class. The data were compared to various models that incorporate a QGP and found to be consistent with the measurement.





\bibliographystyle{elsarticle-num}
\bibliography{sahlmueller_baldo}

\begin{thebibliography}{10}
\expandafter\ifx\csname url\endcsname\relax
  \def\url#1{\texttt{#1}}\fi
\expandafter\ifx\csname urlprefix\endcsname\relax\def\urlprefix{URL }\fi
\expandafter\ifx\csname href\endcsname\relax
  \def\href#1#2{#2} \def\path#1{#1}\fi

\bibitem{Adam:2015lda}
J.~Adam, others (ALICE~Collaboration), {Direct photon production in Pb-Pb
  collisions at $\sqrt{s_{\rm NN}}$ = 2.76 TeV }\href
  {http://arxiv.org/abs/1509.07324} {\path{arXiv:1509.07324}}.

\bibitem{ALICE-PUBLIC-2015-007}
J.~Adam, others (ALICE~Collaboration),
  \href{https://cds.cern.ch/record/2102398}{{Supplemental figures: Direct
  photon production in Pb-Pb collisions at $\sqrt{s_{\rm NN}}$ = 2.76 TeV}}.
\newline\urlprefix\url{https://cds.cern.ch/record/2102398}

\bibitem{Turbide:2003si}
S.~Turbide, R.~Rapp, C.~Gale, {Hadronic production of thermal photons}, Phys.
  Rev. C69 (2004) 014903.
\newblock \href {http://arxiv.org/abs/hep-ph/0308085}
  {\path{arXiv:hep-ph/0308085}}, \href
  {http://dx.doi.org/10.1103/PhysRevC.69.014903}
  {\path{doi:10.1103/PhysRevC.69.014903}}.

\bibitem{Shen:2013vja}
C.~Shen, U.~W. Heinz, J.-F. Paquet, C.~Gale, {Thermal photons as a quark-gluon
  plasma thermometer reexamined}, Phys. Rev. C89~(4) (2014) 044910.
\newblock \href {http://arxiv.org/abs/1308.2440} {\path{arXiv:1308.2440}},
  \href {http://dx.doi.org/10.1103/PhysRevC.89.044910}
  {\path{doi:10.1103/PhysRevC.89.044910}}.

\bibitem{Adare:2008ab}
A.~Adare, others (PHENIX~Collaboration), {Enhanced production of direct photons
  in Au+Au collisions at $\sqrt{s_{NN}}=200$ GeV and implications for the
  initial temperature}, Phys. Rev. Lett. 104 (2010) 132301.
\newblock \href {http://arxiv.org/abs/0804.4168} {\path{arXiv:0804.4168}},
  \href {http://dx.doi.org/10.1103/PhysRevLett.104.132301}
  {\path{doi:10.1103/PhysRevLett.104.132301}}.

\bibitem{Adare:2014fwh}
A.~Adare, others (PHENIX~Collaboration), {Centrality dependence of low-momentum
  direct-photon production in Au$+$Au collisions at $\sqrt{s_{_{NN}}}=200$
  GeV}, Phys. Rev. C91~(6) (2015) 064904.
\newblock \href {http://arxiv.org/abs/1405.3940} {\path{arXiv:1405.3940}},
  \href {http://dx.doi.org/10.1103/PhysRevC.91.064904}
  {\path{doi:10.1103/PhysRevC.91.064904}}.

\bibitem{Aad:2015lcb}
G.~Aad, others (ATLAS~Collaboration), {Centrality, rapidity and transverse
  momentum dependence of isolated prompt photon production in lead-lead
  collisions at $\sqrt{s_{\mathrm{NN}}} = 2.76$ TeV measured with the ATLAS
  detector }\href {http://arxiv.org/abs/1506.08552} {\path{arXiv:1506.08552}}.

\bibitem{Chatrchyan:2012vq}
S.~Chatrchyan, others (CMS~Collaboration), {Measurement of isolated photon
  production in $pp$ and PbPb collisions at $\sqrt{s_{NN}}=2.76$ TeV}, Phys.
  Lett. B710 (2012) 256--277.
\newblock \href {http://arxiv.org/abs/1201.3093} {\path{arXiv:1201.3093}},
  \href {http://dx.doi.org/10.1016/j.physletb.2012.02.077}
  {\path{doi:10.1016/j.physletb.2012.02.077}}.

\bibitem{Paquet:2015lta}
J.-F. Paquet, C.~Shen, G.~S. Denicol, M.~Luzum, B.~Schenke, S.~Jeon, C.~Gale,
  {The production of photons in relativistic heavy-ion collisions}\href
  {http://arxiv.org/abs/1509.06738} {\path{arXiv:1509.06738}}.

\bibitem{vanHees:2014ida}
H.~van Hees, M.~He, R.~Rapp, {Pseudo-Critical Enhancement of Thermal Photons in
  Relativistic Heavy-Ion Collisions}, Nucl. Phys. A933 (2015) 256--271.
\newblock \href {http://arxiv.org/abs/1404.2846} {\path{arXiv:1404.2846}},
  \href {http://dx.doi.org/10.1016/j.nuclphysa.2014.09.009}
  {\path{doi:10.1016/j.nuclphysa.2014.09.009}}.

\bibitem{Chatterjee:2012dn}
R.~Chatterjee, H.~Holopainen, T.~Renk, K.~J. Eskola, {Collision centrality and
  $\tau_0$ dependence of the emission of thermal photons from fluctuating
  initial state in ideal hydrodynamic calculation}, Phys. Rev. C85 (2012)
  064910.
\newblock \href {http://arxiv.org/abs/1204.2249} {\path{arXiv:1204.2249}},
  \href {http://dx.doi.org/10.1103/PhysRevC.85.064910}
  {\path{doi:10.1103/PhysRevC.85.064910}}.

\bibitem{Linnyk:2015tha}
O.~Linnyk, V.~Konchakovski, T.~Steinert, W.~Cassing, E.~L. Bratkovskaya,
  {Hadronic and partonic sources of direct photons in relativistic heavy-ion
  collisions}\href {http://arxiv.org/abs/1504.05699} {\path{arXiv:1504.05699}}.

\end{thebibliography}







\end{document}